\PassOptionsToPackage{unicode}{hyperref}
\PassOptionsToPackage{hyphens}{url}
\PassOptionsToPackage{dvipsnames,svgnames,x11names}{xcolor}
\documentclass[a4paper
]{article}
  \newcommand{\repnuma}{NU-QG-20}
 \newcommand{\repnumb}{RUP-26-12}
 \newcommand{\emaila}{\sf yoo.chulmoon.k6@f.mail.nagoya-u.ac.jp}

\usepackage{fancyhdr}

\fancypagestyle{titlepage}{%
  \fancyhf{}
  \fancyhead[R]{%
    \repnuma\\
     \repnumb
  }
  \fancyhead[L]{%
    \includegraphics[height=2cm]{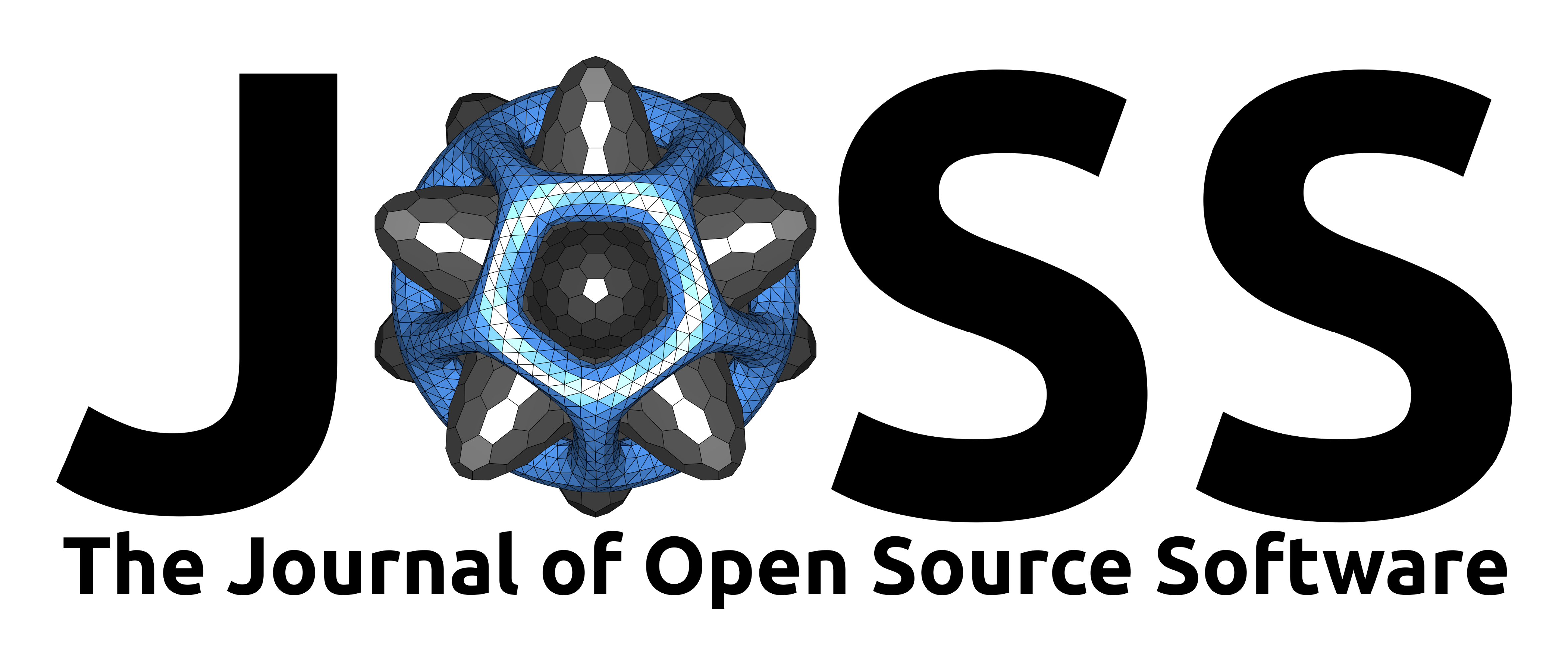}%
  }
   \fancyfoot[L]{%
    \rule{0.2\textwidth}{0.4pt} \\[-\baselineskip]%
    \vspace{4mm}
    \emaila}

}
\makeatletter
\g@addto@macro\maketitle{\thispagestyle{titlepage}} 
\makeatother
\date{}

\usepackage{amsmath,amssymb}
\usepackage{lmodern}
\usepackage{iftex}
\ifPDFTeX
  \usepackage[T1]{fontenc}
  \usepackage[utf8]{inputenc}
  \usepackage{textcomp} 
\else 
  \usepackage{unicode-math}
  \defaultfontfeatures{Scale=MatchLowercase}
  \defaultfontfeatures[\rmfamily]{Ligatures=TeX,Scale=1}
\fi
\IfFileExists{upquote.sty}{\usepackage{upquote}}{}
\IfFileExists{microtype.sty}{
  \usepackage[]{microtype}
  \UseMicrotypeSet[protrusion]{basicmath} 
}{}
\makeatletter
\@ifundefined{KOMAClassName}{
  \IfFileExists{parskip.sty}{%
    \usepackage{parskip}
  }{
    \setlength{\parindent}{0pt}
    \setlength{\parskip}{6pt plus 2pt minus 1pt}}
}{
  \KOMAoptions{parskip=half}}
\makeatother
\usepackage{xcolor}
\usepackage{graphicx}
\makeatletter
\def\maxwidth{\ifdim\Gin@nat@width>\linewidth\linewidth\else\Gin@nat@width\fi}
\def\maxheight{\ifdim\Gin@nat@height>\textheight\textheight\else\Gin@nat@height\fi}
\makeatother
\setkeys{Gin}{width=\maxwidth,height=\maxheight,keepaspectratio}
\makeatletter
\def\fps@figure{ht!}
\makeatother
\NewDocumentCommand\citeproctext{}{}
\NewDocumentCommand\citeproc{mm}{%
  \begingroup\def\citeproctext{#2}\cite{#1}\endgroup}
\makeatletter
 \let\@cite@ofmt\@firstofone
 \def\@biblabel#1{}
 \def\@cite#1#2{{#1\if@tempswa , #2\fi}}
\makeatother
\newlength{\cslhangindent}
\newlength{\csllabelwidth}
\newenvironment{CSLReferences}[2] 
 {\begin{list}{}{%
  \setlength{\itemindent}{0pt}
  \setlength{\leftmargin}{0pt}
  \setlength{\parsep}{0pt}
  \ifodd #1
   \setlength{\leftmargin}{\cslhangindent}
   \setlength{\itemindent}{-1\cslhangindent}
  \fi
  \setlength{\itemsep}{#2\baselineskip}}}
 {\end{list}}
\usepackage{calc}

\ifLuaTeX
\usepackage[bidi=basic]{babel}
\else
\usepackage[bidi=default]{babel}
\fi
\babelprovide[main,import]{american}

\def\languageshorthands#1{}
\ifLuaTeX
  \usepackage{selnolig}  
\fi
\IfFileExists{bookmark.sty}{\usepackage{bookmark}}{\usepackage{hyperref}}
\IfFileExists{xurl.sty}{\usepackage{xurl}}{} 
\hypersetup{
  pdftitle={COSMOS: A numerical relativity code specialized for PBH
formation},
  pdfauthor={Chul-Moon Yoo, Hirotada Okawa, Albert Escrivà, Tomohiro
Harada, Hayami Iizuka, Taishi Ikeda, Yasutaka Koga, Daiki Saito, Masaaki
Shimada, Koichiro Uehara},
  pdflang={en-US},
  colorlinks=true,
  linkcolor={Maroon},
  filecolor={Maroon},
  citecolor={Blue},
  urlcolor={Blue},
  pdfcreator={LaTeX via pandoc}}

\title{COSMOS: A numerical relativity code specialized for PBH formation}

\definecolor{c53baa1}{RGB}{83,186,161}
\definecolor{c202826}{RGB}{32,40,38}

\usepackage[affil-it]{authblk}
\usepackage{orcidlink}
\author[1,2%
  ]{Chul-Moon Yoo%
    \,\orcidlink{0000-0002-9928-4757}\,%
    }
\author[3%
  ]{Hirotada Okawa%
    \,\orcidlink{0000-0001-7372-5131}\,%
    }
\author[1%
  ]{Albert Escriv\`a%
    \,\orcidlink{0000-0001-5483-8034}\,%
    }
\author[4%
  ]{Tomohiro Harada%
    \,\orcidlink{0000-0002-9085-9905}\,%
    }
\author[4%
  ]
  {\hspace{1cm}Hayami Iizuka%
    \,\orcidlink{0009-0001-6604-7763}\,%
    }
\author[5%
  ]{Taishi Ikeda%
    \,\orcidlink{0000-0002-9076-1027}\,%
    }
\author[6%
  ]{Yasutaka Koga%
    \,\orcidlink{0000-0002-9579-5787}\,%
    }
\author[7%
  ]{Daiki Saito%
    \,\orcidlink{0000-0003-1624-9268}\,%
    }
\author[1%
  ]{Masaaki Shimada%
    \,\orcidlink{0009-0001-2144-575X}\,%
    }
\author[1%
  ]{Koichiro Uehara%
    \,\orcidlink{0009-0006-3039-6829}\,%
    }

\affil[1]{\normalsize Graduate School of Science, Nagoya University, Japan%
  }
\affil[2]{\normalsize Kobayashi-Maskawa Institute for the Origin of Particles and the Universe, Nagoya University, Japan%
  }
\affil[3]{\normalsize Faculty of Software and Information Technology, Aomori University, Japan%
  }
\affil[4]{\normalsize Department of Physics, Rikkyo University, Japan%
  }
\affil[5]{\normalsize Center of Gravity, Niels Bohr Institute, Denmark%
  }
\affil[6]{\normalsize Department of Physics, College of Humanities and Sciences,
Nihon University, Japan%
  }
\affil[7]{\normalsize Department of Science Education, Ewha Womans University,
Korea%
  }

\begin{document}
\maketitle

\vspace{-1cm}
DOI: \href{https://doi.org/10.21105/joss.09570}{10.21105/joss.09570}

\section{Summary}\label{summary}

Primordial black holes (PBHs) are black holes generated in the early universe without having gone through stellar evolution. It has been hypothesized that PBHs may be candidates for black holes and compact objects of various masses in the universe or a major component of dark matter. In particular, PBHs have been attracting much attention in the recent development of gravitational wave observation. In the standard
formation process, PBHs are formed from super-horizon primordial fluctuations with non-linearly large initial amplitude. In order to
simulate the non-linear gravitational dynamics of PBH formation, one has
to rely on numerical relativity solvers to approximate the solution of the Einstein equations. \texttt{COSMOS} \footnote{\url{https://github.com/cmyoo/cosmos}} (\citeproc{ref-Okawa:2014nda}{Okawa et al., 2014};
\citeproc{ref-Yoo:2013yea}{Yoo et al., 2013}) provides simple tools for
the simulation of PBH formation (see \texttt{COSMOS-S} \footnote{\url{https://github.com/cmyoo/cosmos-s}} for a spherically symmetric version of \texttt{COSMOS}, which is
not discussed in this paper). \texttt{COSMOS} is a C++ package for
solving the Einstein equations in 3+1 dimensions. It was originally
translated from SACRA code (\citeproc{ref-Yamamoto:2008js}{Yamamoto et
al., 2008}) into C++ and has been developed specifically for the
simulation of PBH formation \footnote{C.Y. and H.O. are the main
  contributors of this code, and other authors used the numerical code
  during the development and operational stages and contributed in part
  to its development and improvement.}. Past publications that use COSMOS for simulation include Yoo et al.
(\citeproc{ref-Yoo:2013yea}{2013}); Okawa et al.
(\citeproc{ref-Okawa:2014nda}{2014}); Yoo \& Okawa
(\citeproc{ref-Yoo:2014boa}{2014}); Okawa \& Cardoso
(\citeproc{ref-Okawa:2014sxa}{2014}); Ikeda et al.
(\citeproc{ref-Ikeda:2015hqa}{2015}); Brito et al.
(\citeproc{ref-Brito:2015yga}{2015}); Brito et al.
(\citeproc{ref-Brito:2015yfh}{2016}); Okawa
(\citeproc{ref-Okawa_2015}{2015}); Yoo et al.
(\citeproc{ref-Yoo:2016kzu}{2017}); Yoo et al.
(\citeproc{ref-Yoo:2018pda}{2019}); Yoo
(\citeproc{ref-Yoo:2024lhp}{2024}); Escrivà \& Yoo
(\citeproc{ref-Escriva:2024lmm}{2025b}); Escrivà \& Yoo
(\citeproc{ref-Escriva:2024aeo}{2025a}) \footnote{In these works,
  additional functions and packages have been implemented that may not
  appear in the public release of COSMOS. Therefore the results may not
  be obtained by simply running the public code.}.


\section{Statement of need}\label{statement-of-need}

In the simulation of PBH formation, the presence of multiple
length scales (the size of the collapsing region and that of cosmological
expansion) necessitates an efficient resolution refinement procedure. In
order to resolve the collapsing region, non-Cartesian scale-up coordinates (\citeproc{ref-Yoo:2018pda}{Yoo et al., 2019}) and a fixed
mesh-refinement procedure (\citeproc{ref-Yoo:2024lhp}{Yoo, 2024}) are
implemented in \texttt{COSMOS}. In its model, COSMOS uses a perfect
fluid with a linear equation of state and a massless scalar field as
matter fields. To achieve a practically acceptable computational speed,
OpenMP is used for the parallelization. COSMOS has no other
dependencies, which makes for an easier installation. Once users
understand the source code to some extent, the system can be easily
extended to various scientifically interesting settings.

\section{Physical system settings}\label{physical-system-settings}

At the core of COSMOS, the Einstein equations \[
G_{\mu\nu}=R_{\mu\nu}-\frac{1}{2}Rg_{\mu\nu}=\frac{8\pi G}{c^4}T_{\mu\nu}
\] are solved, where \(G_{\mu\nu}\), \(g_{\mu\nu}\), \(R_{\mu\nu}\),
\(R\), \(G\), \(c\) and \(T_{\mu\nu}\) are the Einstein tensor, the
metric tensor, the Ricci tensor, the Ricci scalar, the Newtonian
gravitational constant, the speed of light, and the energy-momentum
tensor, respectively. The energy-momentum tensor is divided into fluid
and scalar field contributions as \[
T_{\mu\nu}=T^{\rm SC}_{\mu\nu}+T^{\rm FL}_{\mu\nu}, 
\] with \[
T^{\rm SC}_{\mu\nu}=\nabla_\mu\phi\nabla_\nu\phi-\frac{1}{2}g_{\mu\nu}\nabla^\lambda\phi\nabla_\lambda\phi 
\] and \[
T^{\rm FL}_{\mu\nu}=(\rho+P)u_\mu u_\nu+Pg_{\mu\nu}, 
\] where \(\nabla\) denotes taking the covariant derivative using
\(g_{\mu\nu}\), \(\phi\) is the scalar field, and \(\rho\), \(u_\mu\)
and \(P\) are the energy density, the four-velocity, and the pressure of
the fluid, respectively. The pressure and the energy density are assumed
to satisfy the linear equation of state \(P=w\rho\) with \(w\) being a
constant. The equations of motion for the scalar field \[
\nabla^\mu\nabla_\mu \phi=0, 
\] and the fluid \[
\nabla^\mu T^{\rm FL}_{\mu\nu}=0, 
\] are also solved. Readers are asked to refer to standard textbooks of
numerical relativity (e.g., Gourgoulhon
(\citeproc{ref-gourgoulhon20123+1}{2012}); Shibata
(\citeproc{ref-shibata2016numerical}{2016})) on how to rewrite these
equations into a form suitable for numerical integration. To solve the
fluid equations of motion, we basically follow the scheme discussed in
Kurganov \& Tadmor (\citeproc{ref-Kurganov:2000ovy}{2000}); Shibata \&
Font (\citeproc{ref-Shibata:2005jv}{2005}).

As for the initial data, we adopt the long-wavelength growing-mode
solutions up to (and including) the next-leading order of the expansion
parameter \(\epsilon=k/(aH)\ll1\), where \(1/k\) gives the
characteristic comoving scale of the inhomogeneity, and \(a\) and \(H\)
are the scale factor and Hubble expansion rate in the reference
background universe. The initial data can be characterized by a function
of the spatial coordinates \(\vec x\) as the curvature perturbation
\(\zeta(\vec x)\) for adiabatic fluctuations
(\citeproc{ref-Harada:2015yda}{Harada et al., 2015};
\citeproc{ref-Yoo:2020lmg}{Yoo et al., 2020};
\citeproc{ref-Yoo:2024lhp}{Yoo, 2024}) and iso-curvature perturbation
\(\Upsilon(\vec x)\) for massless scalar iso-curvature
(\citeproc{ref-Yoo:2021fxs}{Yoo et al., 2022}). Since the space is
filled with the fluid, the initial fluid distribution can be generated
to meet the constraint equations included in the Einstein equations.
Therefore, the constraint equations are initially satisfied to within machine precision, and need not be solved by integrating elliptic differential equations. This approach differs from the standard method of obtaining the initial data for spacetimes with asymptotically flat vacuum regions, and is the reason why elliptic solvers are not included
in COSMOS.

\section{Examples}\label{examples}

Three examples are included in the package. These examples are intended
primarily for demonstration and instructional purposes, and thus they
are not highly accurate. The resolution has been intentionally kept as
low as possible. In the figures below, the length scale is normalized by
the size \(L\) of the box for the numerical simulation.

\subsubsection{Evolution of a single-mode
perturbation}\label{evolution-of-a-single-mode-perturbation}

\begin{figure}
\centering
\includegraphics[width=\linewidth,height=6cm,keepaspectratio]{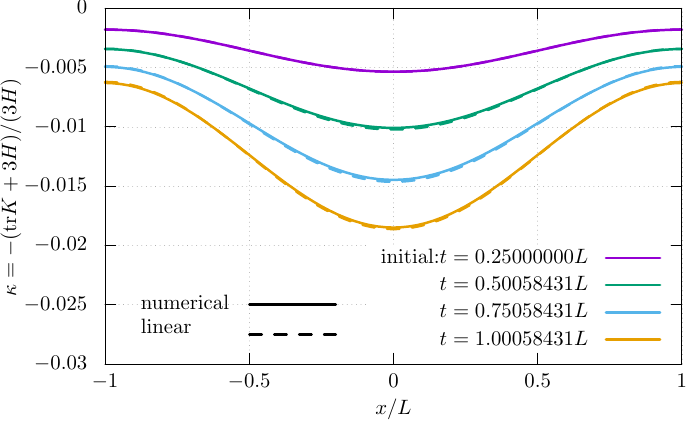}
\caption{The time evolution of the trace of the extrinsic curvature
tr\(K\) is compared with the solution of the linear perturbation
equation.\label{fig:kap}}
\end{figure}

The evolution of a small sinusoidal fluctuation is given as an example,
which can be compared to the corresponding linear perturbation (see
\autoref{fig:kap}).

\subsubsection{Adiabatic spherically symmetric initial
fluctuation}\label{adiabatic-spherically-symmetric-initial-fluctuation}

\begin{figure}
\centering
\includegraphics[width=\linewidth,height=7cm,keepaspectratio]{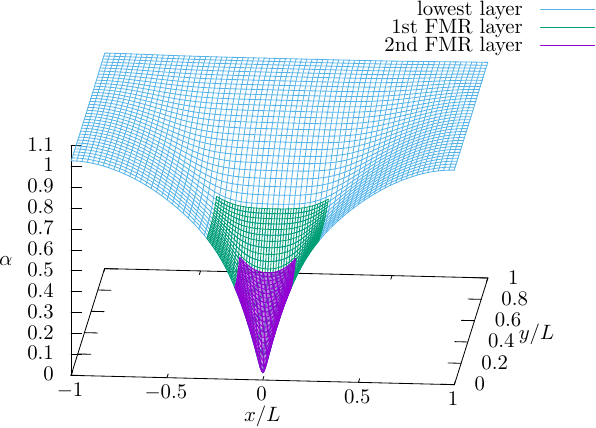}
\caption{The lapse function (``\(tt\)-component'' of the metric) on the
\(xy\)-plane at the time when an apparent horizon is found. The blue,
green, and purple meshes show the region covered by the lowest, 1st, and
2nd mesh refinement layers, respectively.\label{fig:alp}}
\end{figure}

\begin{figure}
\centering
\includegraphics[width=\linewidth,height=5cm,keepaspectratio]{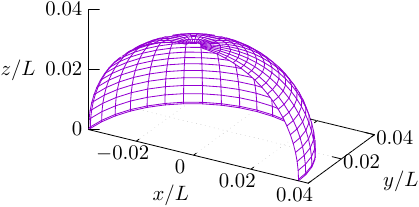}
\caption{The shape of the apparent horizon when it is
found.\label{fig:AH}}
\end{figure}

Here, we consider the adiabatic perturbation generated by the initial
curvature perturbation without scalar field contribution. As is
described in Harada et al. (\citeproc{ref-Harada:2015yda}{2015}), once
the spatial profile of the curvature perturbation \(\zeta(\vec x)\) is
specified, the growing mode solution can be described in the long-wavelength approximation. 
More details, including the specific functional form of the curvature perturbation, can be found in the instruction page\footnote{\url{https://github.com/cmyoo/cosmos/wiki/Adiabatic-spherically-symmetric-initial-fluctuation}} (see also Yoo et al. (\citeproc{ref-Yoo:2020lmg}{2020})). 
In the repository, we include the data file \texttt{ini\_all.dat} necessary to reconstruct the geometry and matter distribution at the time an apparent
horizon is found (see \autoref{fig:alp} and \autoref{fig:AH}, which can be generated by following the instructions).

\subsubsection{Spherically symmetric
iso-curvature}\label{spherically-symmetric-iso-curvature}

Here, we consider the iso-curvature perturbation generated by a massless
scalar field. We assume that the massless scalar field does not
contribute to the background metric of the gradient expansion. Then, as
is described in Yoo et al. (\citeproc{ref-Yoo:2021fxs}{2022}), once the
spatial profile of the iso-curvature mode \(\Upsilon(\vec x)\) is
specified at the leading order of the gradient expansion, the growing mode solution can be described in the long-wavelength approximation.
More details, including the specific functional form of \(\Upsilon(\vec x)\) can be found in the instruction page\footnote{\url{https://github.com/cmyoo/cosmos/wiki/Spherically-symmetric-isocurvature}} (see also Yoo et al. (\citeproc{ref-Yoo:2021fxs}{2022})). In the repository, we include the data file \texttt{ini\_all.dat} necessary to reconstruct the geometry and matter distribution at the time an apparent horizon is found, as for the adiabatic case.

\section{Acknowledgements}\label{acknowledgements}

A.E. acknowledges support from the JSPS Postdoctoral Fellowships for
Research in Japan (Graduate School of Sciences, Nagoya University). K.U.
would like to take this opportunity to thank the ``THERS Make New Standards Program for the Next Generation Researchers'' supported by JST
SPRING, Grant Number JPMJSP2125. T.I. acknowledges support from VILLUM
Foundation (grant no. VIL37766) and the DNRF Chair program (grant no.
DNRF162) by the Danish National Research Foundation. This work is
supported in part by JSPS KAKENHI Grant Nos. 20H05850 (C.Y.), 20H05853
(T.H., C.Y.), 21K20367 (Y.K.), 23KK0048 (Y.K.), 24K07027 (T.H., C.Y.), 24KJ1223 (D.S.), and 25K07281 (C.Y.).

\section*{References}\label{references}
\addcontentsline{toc}{section}{References}

\protect\phantomsection\label{refs}


\begin{CSLReferences}{1}{0}
\bibitem[\citeproctext]{ref-Brito:2015yfh}
Brito, R., Cardoso, V., Macedo, C. F. B., Okawa, H., \& Palenzuela, C.
 {\em Interaction between bosonic dark matter and stars}. {Phys. Rev. D}, {\bf 93}, 044045 (2016).
\url{https://doi.org/10.1103/PhysRevD.93.044045}

\bibitem[\citeproctext]{ref-Brito:2015yga}
Brito, R., Cardoso, V., \& Okawa, H.  {\em Accretion of dark matter
by stars}. {Phys. Rev. Lett.}, {\bf 115}, 111301 (2015).
\url{https://doi.org/10.1103/PhysRevLett.115.111301}

\bibitem[\citeproctext]{ref-Escriva:2024aeo}
Escrivà, A., \& Yoo, C.-M. {\em Nonspherical effects on the mass
function of primordial black holes}. {Phys. Rev. D}, {\bf 112},
L081304 (2025). \url{https://doi.org/10.1103/4jbp-87wc}

\bibitem[\citeproctext]{ref-Escriva:2024lmm}
Escrivà, A., \& Yoo, C.-M.  {\em Simulations of ellipsoidal
primordial black hole formation}. {Phys. Rev. D}, {\bf 112},
083518 (2025).  \url{https://doi.org/10.1103/PhysRevD.112.083518}

\bibitem[\citeproctext]{ref-gourgoulhon20123+1}
Gourgoulhon, E. \emph{\em 3+1 formalism in general relativity :
Bases of numerical relativity}. Springer  (2012).
\url{https://doi.org/10.1007/978-3-642-24525-1}

\bibitem[\citeproctext]{ref-Harada:2015yda}
Harada, T., Yoo, C.-M., Nakama, T., \& Koga, Y. {\em Cosmological
long-wavelength solutions and primordial black hole formation}.
{Phys. Rev. D}, {\bf 91}, 084057 (2015).
\url{https://doi.org/10.1103/PhysRevD.91.084057}

\bibitem[\citeproctext]{ref-Ikeda:2015hqa}
Ikeda, T., Yoo, C.-M., \& Nambu, Y. {\em Expanding universe with
nonlinear gravitational waves}. {Phys. Rev. D}, {\bf 92},
044041 (2015). \url{https://doi.org/10.1103/PhysRevD.92.044041}

\bibitem[\citeproctext]{ref-Kurganov:2000ovy}
Kurganov, A., \& Tadmor, E. {\em New high-resolution central schemes
for nonlinear conservation laws and convection\textendash{}diffusion
equations}. {J. Comput. Phys.}, {\bf 160}, 241--282 (2000). 
\url{https://doi.org/10.1006/jcph.2000.6459}

\bibitem[\citeproctext]{ref-Okawa_2015}
Okawa, H. {\em Nonlinear evolutions of bosonic clouds around black
holes}. {Classical and Quantum Gravity}, {\bf 32}, 214003 (2015).
\url{https://doi.org/10.1088/0264-9381/32/21/214003}

\bibitem[\citeproctext]{ref-Okawa:2014sxa}
Okawa, H., \& Cardoso, V. {\em Black holes and fundamental fields:
Hair, kicks, and a gravitational Magnus effect}. {Phys. Rev. D},
\emph{\bf 90}, 104040 (2014). \url{https://doi.org/10.1103/PhysRevD.90.104040}

\bibitem[\citeproctext]{ref-Okawa:2014nda}
Okawa, H., Witek, H., \& Cardoso, V. {\em Black holes and
fundamental fields in Numerical Relativity: initial data construction
and evolution of bound states}. {Phys. Rev. D}, {\bf 89},
104032 (2014). \url{https://doi.org/10.1103/PhysRevD.89.104032}

\bibitem[\citeproctext]{ref-shibata2016numerical}
Shibata, M. \emph{Numerical relativity}. World Scientific
Publishing Co. Pte. Ltd. (2016). \url{https://doi.org/10.1142/9692}

\bibitem[\citeproctext]{ref-Shibata:2005jv}
Shibata, M., \& Font, J. A. {\em Robustness of a high-resolution
central scheme for hydrodynamic simulations in full general relativity}. {Phys. Rev. D}, {\bf 72}, 047501 (2005). 
\url{https://doi.org/10.1103/PhysRevD.72.047501}

\bibitem[\citeproctext]{ref-Yamamoto:2008js}
Yamamoto, T., Shibata, M., \& Taniguchi, K. {\em Simulating
coalescing compact binaries by a new code SACRA}. {Phys. Rev. D}, {\bf 78}, 064054 (2008). \url{https://doi.org/10.1103/PhysRevD.78.064054}

\bibitem[\citeproctext]{ref-Yoo:2024lhp}
Yoo, C.-M. {\em Primordial black hole formation from a nonspherical
density profile with a misaligned deformation tensor}. {Phys. Rev. D}, {\bf 110}, 043526 (2024). 
\url{https://doi.org/10.1103/PhysRevD.110.043526}

\bibitem[\citeproctext]{ref-Yoo:2021fxs}
Yoo, C.-M., Harada, T., Hirano, S., Okawa, H., \& Sasaki, M. 
{\em Primordial black hole formation from massless scalar isocurvature}.
{Phys. Rev. D}, {\bf 105}, 103538 (2022). 
\url{https://doi.org/10.1103/PhysRevD.105.103538}

\bibitem[\citeproctext]{ref-Yoo:2016kzu}
Yoo, C.-M., Harada, T., \& Okawa, H. {\em 3D Simulation of spindle
gravitational collapse of a collisionless particle system}. {Class. Quant. Grav.}, {\bf 34}, 105010 (2017). 
\url{https://doi.org/10.1088/1361-6382/aa6ad5}

\bibitem[\citeproctext]{ref-Yoo:2020lmg}
Yoo, C.-M., Harada, T., \& Okawa, H. {\em Threshold of primordial black hole formation in nonspherical collapse}. {Phys. Rev. D},
{\bf 102}(4), 043526 (2020). \url{https://doi.org/10.1103/PhysRevD.102.043526}

\bibitem[\citeproctext]{ref-Yoo:2018pda}
Yoo, C.-M., Ikeda, T., \& Okawa, H. {\em Gravitational collapse of a
massless scalar field in a periodic box}. {Class. Quant. Grav.},
{\bf 36}, 075004 (2019).  \url{https://doi.org/10.1088/1361-6382/ab06e2}

\bibitem[\citeproctext]{ref-Yoo:2014boa}
Yoo, C.-M., \& Okawa, H. {\em Black hole universe with a
cosmological constant}. {Phys. Rev. D}, {\bf 89}, 123502 (2014). 
\url{https://doi.org/10.1103/PhysRevD.89.123502}

\bibitem[\citeproctext]{ref-Yoo:2013yea}
Yoo, C.-M., Okawa, H., \& Nakao, K. {\em Black Hole Universe: Time
Evolution}. {Phys. Rev. Lett.}, {\bf 111}, 161102 (2013). 
\url{https://doi.org/10.1103/PhysRevLett.111.161102}

\end{CSLReferences}
\end{document}